\documentclass[modern]{aastex61}

\usepackage{amsmath,amstext}
\usepackage[T1]{fontenc}
\usepackage{apjfonts} 
\usepackage{hyperref}
\usepackage[figure,figure*]{hypcap}
\hypersetup{colorlinks, citecolor=blue, filecolor=blue, linkcolor=blue, urlcolor=blue}

\shorttitle{Plasma evolution within an erupting coronal cavity}
\shortauthors{Long et al.}
\newcommand{\kms}{km~s$^{-1}$}
\newcommand{\eg}{\emph{e.g.}}
\newcommand{\cf}{\emph{cf.}}
\newcommand{\ie}{\emph{i.e.}}
\makeatletter

\newcommand{\Rmnum}[1]{\expandafter\@slowromancap\romannumeral #1@}
\makeatother
\newcommand{\corr}[1]{\textcolor{black}{ #1}}
\newcommand{\ctwo}[1]{\textcolor{black}{ #1}}

\received{}
\revised{ }
\accepted{ }

\submitjournal{ApJ}

\begin{document}

\title{Plasma evolution within an erupting coronal cavity}

\correspondingauthor{David~M.~Long}
\email{david.long@ucl.ac.uk}

\author[0000-0003-3137-0277]{David~M.~Long}
\author[0000-0001-9457-6200]{Louise~K.~Harra}
\author[0000-0001-9346-8179]{Sarah~A.~Matthews}
\affil{UCL-Mullard Space Science Laboratory, Holmbury St.~Mary, Dorking, Surrey, RH5~6NT, UK}

\author[0000-0001-6102-6851]{Harry~P.~Warren}
\affil{Space Science Division, Naval Research Laboratory, Washington, DC 20375, USA}

\author{Kyoung-Sun~Lee}
\affil{National Astronomical Observatory of Japan (NAOJ), 2-21-1, Osawa, Mitaka, Tokyo 181-8588}

\author{George~A.~Doschek}
\affil{Space Science Division, Naval Research Laboratory, Washington, DC 20375, USA}

\author{Hirohisa~Hara}
\affil{National Astronomical Observatory of Japan (NAOJ), 2-21-1, Osawa, Mitaka, Tokyo 181-8588}

\author[0000-0002-8975-812X]{Jack~M.~Jenkins}
\affil{UCL-Mullard Space Science Laboratory, Holmbury St.~Mary, Dorking, Surrey, RH5~6NT, UK}

\begin{abstract}
Coronal cavities have previously been observed associated with long-lived quiescent filaments and are thought to correspond 
to the associated magnetic flux rope. Although the standard flare model predicts a coronal cavity corresponding to the 
erupting flux rope, these have only been observed using broadband imaging data, restricting analysis to the plane-of-sky. We 
present \corr{a unique set of} spectroscopic observations of an active region filament seen \corr{erupting at the solar limb} 
in the extreme ultraviolet (EUV). The cavity erupted and expanded rapidly, with the change in rise phase contemporaneous with 
an increase in non-thermal electron energy flux of the associated flare. Hot and cool filamentary material was observed to 
rise with the erupting flux rope, disappearing suddenly as the cavity appeared. Although strongly blue-shifted plasma continued 
to be observed flowing from the apex of the erupting flux rope, this outflow soon ceased. These results indicate that the 
sudden injection of energy from the flare beneath forced the rapid eruption and expansion of the flux rope, driving strong 
plasma flows which resulted in the eruption of an under-dense filamentary flux rope.
\end{abstract}


\section{Introduction}
\label{s:intro}
Solar eruptions are the most energetic and spectacular events that occur in the solar system. However, the processes 
leading to their initiation and how they subsequently evolve remain areas of interest to the solar physics community. 
The currently accepted model of a solar eruption (commonly called the `standard flare model') was originally proposed 
by \citet{Carmichael:1964,Sturrock:1966,Hirayama:1974} and \citet{Kopp:1976}. This model describes a solar flare as a 
brightening driven by magnetic reconnection of coronal loops during the eruption of a magnetic flux rope which is 
subsequently observed in the corona as a coronal mass ejection (CME). While this model has begun to be supplanted 
by more physically realistic, 3-dimensional interpretations \citep[\eg][]{Janvier:2014}, the basic configuration 
remains the same. The origin of erupting magnetic flux ropes also continues to be a source of investigation, with a 
debate as whether they are pre-existing magnetic structures or are formed ``on-the-fly'' during an eruption \citep[see, 
\eg\ the papers by][]{Forbes:2000,Chen:2011,Patsourakos:2013}.

As magnetic structures in the solar corona, flux ropes are difficult to observe directly. Their involvement in eruptions is 
typically inferred by either extrapolating the pre-eruption photospheric magnetic field or via direct measurement of the 
magnetic field of the associated CME \emph{in-situ} following eruption. However, the existence of a pre-eruptive flux rope 
configuration can also be inferred through the observations of dark cavities in white light, extreme ultraviolet (EUV) and 
X-ray observations when located at or near the solar limb 
\citep[\cf][]{Gibson:2006,Habbal:2010,Kucera:2012,Kathy:2012,Karna:2015}. These cavities are generally associated with 
quiescent prominences at the limb, leading to the conclusion that the cavity is a magnetic flux rope supporting cool 
filamentary material in the lower apex \citep[\cf][]{Regnier:2011,Berger:2012,Forland:2013}. Although long-lived, these 
cavity structures can ultimately become unstable and erupt, producing CMEs observable in white-light coronagraph data 
\citep[\eg][]{Sterling:2004,Gibson:2006}. 

While numerous observations exist of coronal cavities associated with quiescent filaments, coronal cavities associated 
with eruptive events are much rarer \ctwo{\citep[although some observations have been presented by, \eg][]{Kleint:2015,Mccauley:2015}}. Limb 
observations of eruptive events, particularly using emission measure analysis techniques, have revealed magnetic flux ropes 
with a hot, bright sheath surrounding a cool, dark cavity which contains a hot, bright core 
\citep[\cf][]{Hannah:2013,Kumar:2014,Lee:2017}. These observations have also been used to try and probe the characteristics 
of the erupting plasma, allowing the density and plane-of-sky velocity of the plasma to be estimated. \citet{Hannah:2013} 
also found that the kinetic energy of the core and current sheet continued to grow during the eruption, indicating that there 
was a continuous input of energy. However, spectroscopic observations of erupting coronal cavities, which provide valuable 
information on how the plasma contained within these cavities is moving in 3-dimensions, remain frustratingly rare. This is 
primarily due to the very small field-of-view of the available instruments and the long time-scales required to obtain the 
observations, both of which require precise advance knowledge of the location of the eruption. 

In this paper, we report on a unique set of spectroscopic observations of a large eruptive event with an associated coronal 
cavity across a broad temperature range. This is the first time that such a cavity eruption has been observed spectroscopically 
\corr{erupting from an active region}, with the observations providing a unique physical insight into this phenomenon.

\section{Observations and Data Analysis}
\label{s:obs}

\begin{figure*}[!ht]
\centering
\includegraphics[width = 0.95\textwidth, trim=0 5 0 7,clip=]{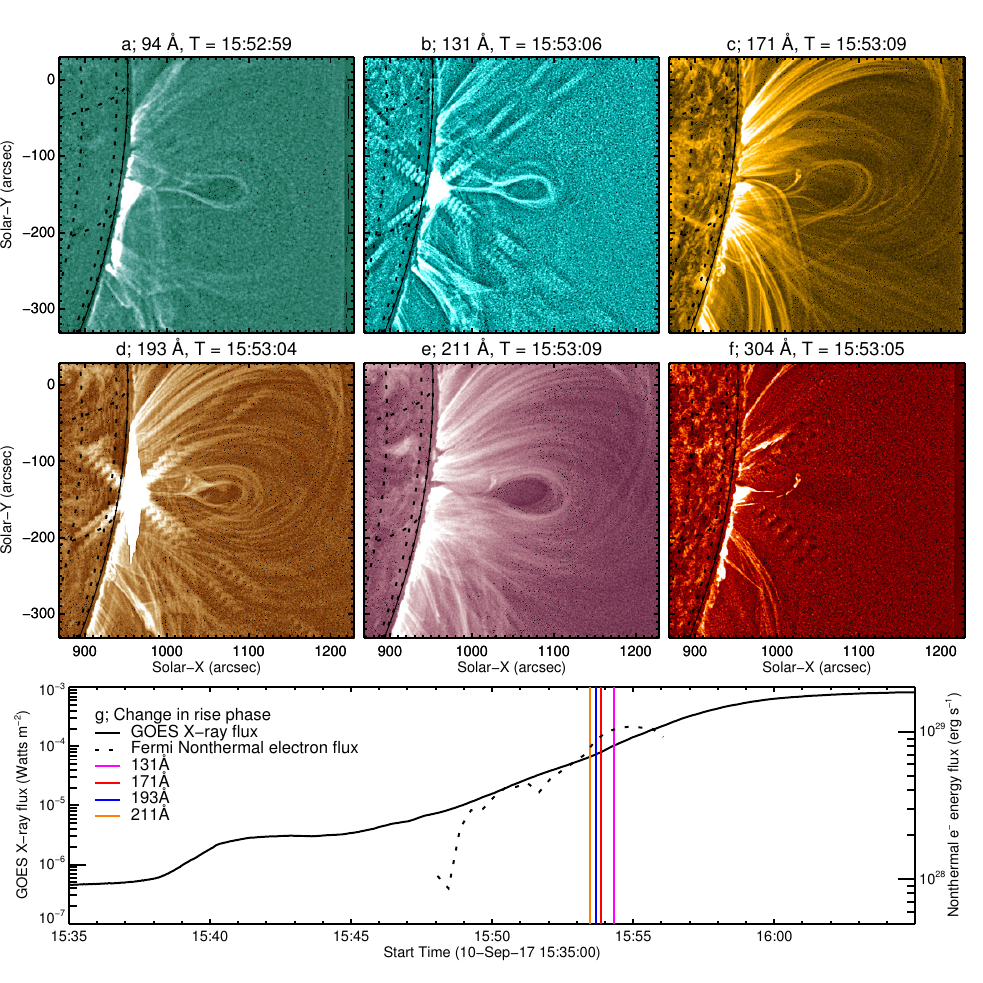}
\caption{The eruption from 2017-Sept-10 at 15:53:06~UT as observed by \emph{SDO}/AIA in the 94~\AA\ (panel~a), 
		 131~\AA\ (panel~b), 171~\AA\ (panel~c), 193~\AA\ (panel~d), 211~\AA\ (panel~e) and 304~\AA\ (panel~f) 
         passbands. Each image has been processed using the Multiscale Gaussian Normalisation (MGN) technique 
         of \citet{Morgan:2014} to highlight the fine structure. \emph{Panel~g}; GOES X-ray flux showing the 
         flare (solid black line) and nonthermal electron energy flux from \emph{Fermi} (dashed black line). 
         Vertical coloured lines show when the rise phase of the erupting cavity changed in each passband from 
         slow to fast (see Figure~\ref{fig:cavity_height}). Note that the diagonal patterns in panels~b, d and 
         f emanating from the site of the flare are diffraction patterns and are unphysical.}
\label{fig:context}
\end{figure*}

The event studied here erupted from the west limb of the Sun on 2017~September~10 and was associated with a GOES 
X8.2 class flare which began at 15:35~UT and peaked at 16:06~UT. The eruption came from NOAA active region AR~12673, 
and was one of a series of major flares produced by the active region following emergence. The evolution of the active 
region and the eruption discussed here were well observed by the Atmospheric Imaging Assembly \citep[AIA;][]{Lemen:2012} 
onboard the \emph{Solar Dynamics Observatory} \citep[SDO;][]{Pesnell:2012} spacecraft. However, AR~12673 was a highly 
energetic active region and as a result was also well observed by multiple space-based and ground-based instruments as it 
evolved and transited the solar limb. 

Following a series of major flares, the active region became the focus of a major flare watch campaign, with regular observations 
from the \emph{Extreme ultraviolet Imaging Spectrometer} \citep[EIS;][]{Culhane:2007} onboard the \emph{Hinode} spacecraft 
\citep{Kosugi:2007}. The primary observing plan involved a scanning raster campaign designed to study post-eruption supra-arcade 
plasma (\emph{Hinode} Observing Plan 244). This campaign, which used EIS study \emph{FlareResponse01}, rastered the 2\arcsec\ 
slit from right to left across a field of view of $239\arcsec \times 304\arcsec$, observing a range of emission lines from 
He~\Rmnum{2} (at log~T=4.7) to Fe~\Rmnum{24} (at log~T=7.2). As the active region transited the west limb, the campaign was 
designed to begin off-limb and raster towards the solar disk, taking $\approx$8~minutes~52~seconds to complete each raster. As 
a result, the erupting cavity described here was only observed by the rasters which began at 15:42:26~UT and 15:51:18~UT 
respectively, with both rasters used in this analysis. Before analysis, the EIS data in each case were first aligned to the 
SDO/AIA field-of-view to allow a direct comparison between the data-sets used. 

\begin{figure*}[!ht]
\centering
\includegraphics[width = 0.99\textwidth, trim=0 0 15 0,clip=]{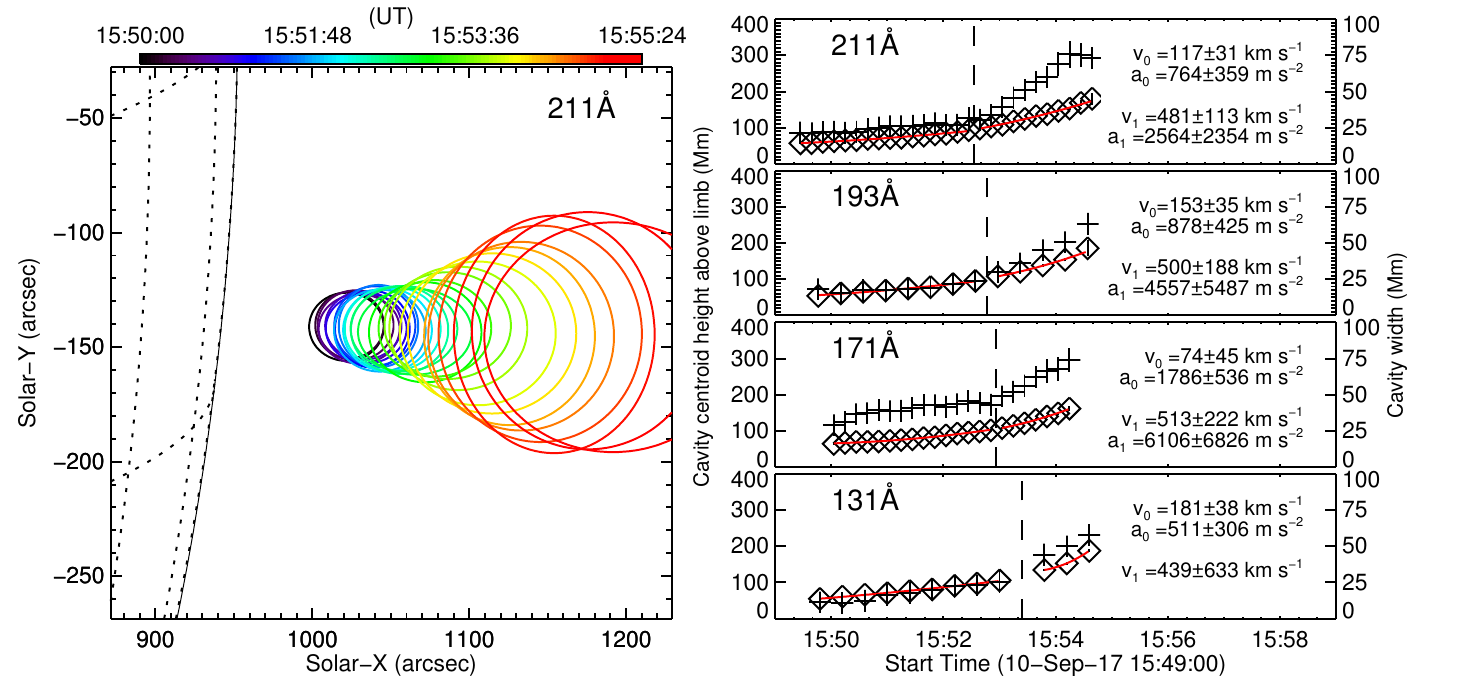}
\caption{\emph{Left panel}; Ellipses fitted to the evolving cavity identified using the 211~\AA\ passband. \emph{Right column}; 
		 Temporal evolution of the height above the limb of the \corr{centroid of the ellipse fitted to the cavity} (diamonds) 
         and \corr{full length of the minor axis of the ellipse fitted to} the cavity (crosses) in each passband studied. The 
         height-time evolution was best fitted using two separate quadratic functions consistent with two distinct rise phases 
         (\ie, a slow then fast rise phase). The point at which the rise phase changed is indicated by the vertical dashed line 
         and was identified using the variation in ellipse width. Note that the separation of the vertical dashed line 
         corresponding to the 131~\AA\ passband and the missing acceleration value for the second rise phase are due to data gaps 
         as a result of images affected by the Automatic Exposure Control being ignored.}
\label{fig:cavity_height}
\end{figure*}

These narrowband spectroscopic observations from \emph{Hinode}/EIS were complemented using the broadband full-Sun 
observations \corr{from SDO/AIA}. This allowed the erupting feature to be tracked at high cadence using multiple 
passbands at multiple temperatures, with the 131~\AA, 171~\AA, 193~\AA\ and 211~\AA\ passbands providing the clearest 
observations of the erupting cavity feature. The data were processed using the standard \emph{aia\_prep.pro} routine in 
SolarSoftWare, with the data also deconvolved with the relevant point spread functions using the 
\emph{aia\_deconvolve\_richardsonlucy.pro} routine for the differential emission measure analysis outlined in Section~\ref{s:dem}.

\section{Results}\label{s:res}

A snapshot of the erupting cavity as observed by the six EUV passbands onboard SDO/AIA at T$\sim$15:53:06~UT is shown in 
Figure~\ref{fig:context}. It is clear (particularly from the 94~\AA, 131~\AA, 171~\AA, 193~\AA\ and 211~\AA\ passbands 
in panels~a--e respectively) that the eruption appears to be a textbook example of the ``standard flare model'', with a 
thin bright feature (consistent with a current sheet) connecting the bright flare loops below to the dim cavity above. Note 
that these images have been processed using the Multiscale Gaussian Normalisation (MGN) technique of \citet{Morgan:2014} 
to highlight the fine structure of the erupting cavity and the associated current sheet. Although both features were observed by 
\emph{Hinode}/EIS, only the cavity is discussed here, with the evolution of the current sheet discussed by \citet{Warren:2017}. 
While the cavity is less clear in the 304~\AA\ passband image shown in panel~f, a thin bright curved structure corresponding to 
the filamentary plasma is visible at the lower apex of the teardrop-shaped cavity. The evolution of this material is clear in the 
movie attached to Figure~\ref{fig:context}; as the cavity forms and rises, bright material can be seen rising with it at the lower 
apex of the teardrop. This material then disappears in all passbands almost simultaneously at $\sim$15:51:30~UT. 

The temporal evolution of the cavity was tracked using the 131~\AA, 171~\AA, 193~\AA\ and 211~\AA\ passbands by fitting an 
ellipse to the bright edge of the cavity at each time step in each passband. Although this is shown in 
Figure~\ref{fig:cavity_height}a solely for the 211~\AA\ passband, the evolution of the cavity was comparable in each passband 
studied. In each case the data were processed using the MGN technique, with the edge of the cavity then manually identified 
and fitted using a ellipse. The temporal evolution of both the centroid (denoted by the diamonds) and width (denoted by the 
cross) of the fitted ellipses are shown in the right column of Figure~\ref{fig:cavity_height} for each passband. A similar 
evolution was observed in each passband, with the cavity found to exhibit two distinct rise phases, particularly when considering 
the evolution of cavity width. The vertical dashed lines in each of the panels on the right-hand column of 
Figure~\ref{fig:cavity_height} indicate the point at which the change in rise phase occurs in each passband. The data were best 
fitted using two independent quadratic functions, with the resulting fit parameters given in the right-hand side of each 
panel. These measurements indicate that the cavity initially rose slowly then more rapidly; an observation consistent with 
previous analyses of the initial stages of solar eruptive events \corr{\citep[\cf][]{Regnier:2011,Byrne:2014}}. 

\begin{figure*}[!ht]
\centering
\includegraphics[width = \textwidth, trim=0 0 0 0,clip=]{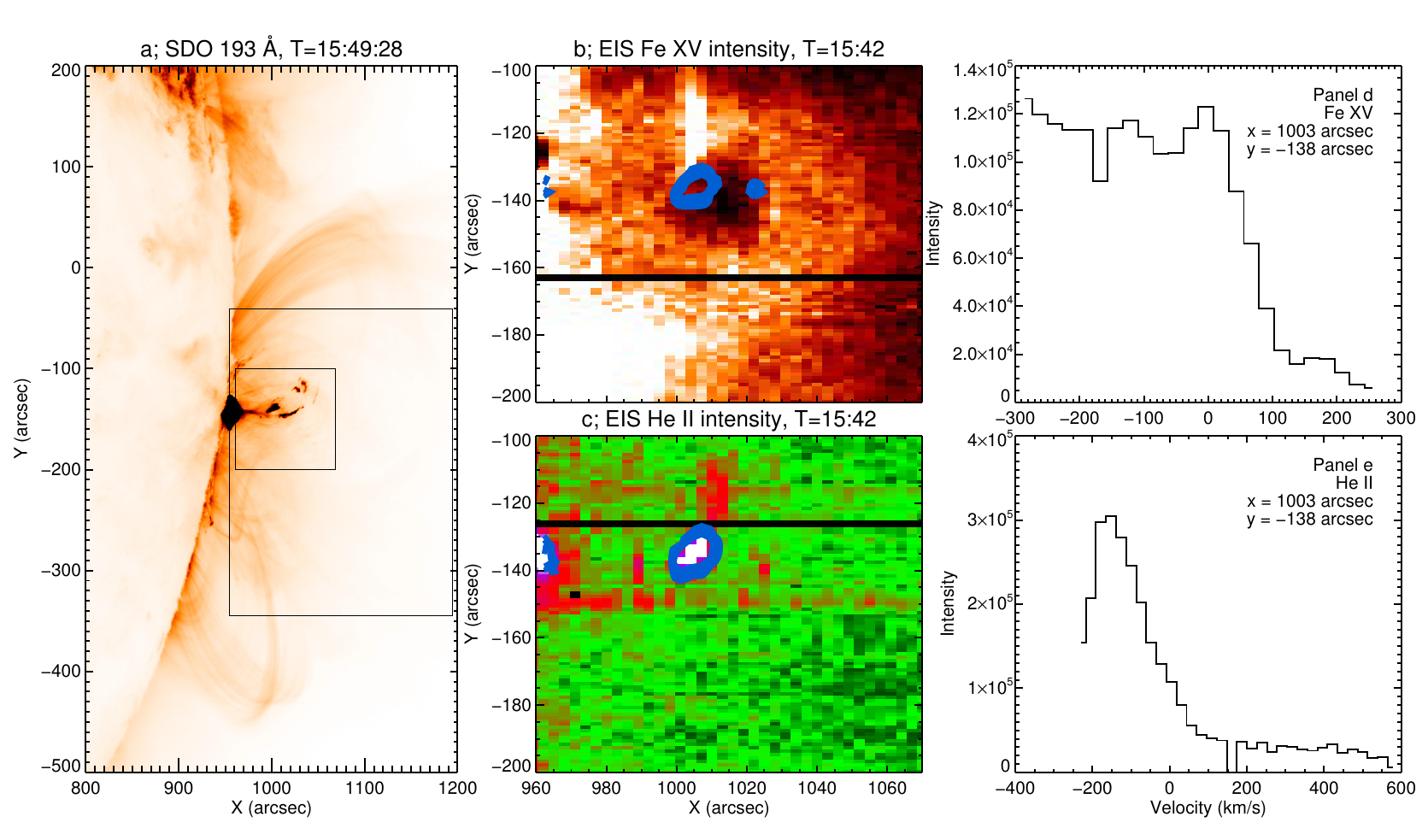}
\caption{\emph{Panel a}; SDO/AIA 193~\AA\ image at 15:49:28~UT showing the full \emph{Hinode}/EIS field-of-view (large 
		 solid box) and the field of view used in panels~b \& c (small solid box). \emph{Panel~b}; Fe~\Rmnum{15} 
         intensity image from \emph{Hinode}/EIS. \emph{Panel~c}; as panel~b, but for the He~\Rmnum{2} emission line. 
         Blue contours in both panels~b \& c indicate where the plasma is blue-shifted by 214~\kms. Note that the cavity 
         is most apparent in the Fe~\Rmnum{15} image (panel~b), while the filamentary material is most apparent in the 
         He~\Rmnum{2} image (panel~c). \emph{Panels~d} \& \emph{e} show sample Doppler-shifted spectra from the 
         Fe~\Rmnum{15} and He~\Rmnum{2} emission lines respectively, illustrating their complex nature. \corr{These 
         Doppler-shifted spectra were constructed by converting the spectrum in the pixel defined in the legend from 
         wavelength to velocity space using a rest wavelength estimated by averaging the spectra in a nearby portion of 
         quiet activity.}}
\label{fig:EIS1542}
\end{figure*}

The points at which the rise phase changed in each passband are also shown in panel~g of Figure~\ref{fig:context}, which shows 
the evolution of the soft X-ray flux from the GOES spacecraft and the non-thermal electron energy flux derived from the Fermi 
large-area telescope \citep[LAT;][]{Atwood:2009}. Although the flare was seen by all of the Fermi/LAT detectors as a result of 
its whole Sun field-of-view, only the NAL04 detector was used here as it showed the least saturation at the lower energies. The 
Fermi spectra were accumulated over 24~s (since NAL04 was not the most sunward facing detector) and fitted with a combination 
of a thermal + thick target function over the energy range 10--100~kev. The nonthermal energy was then calculated at that 
cadence. It is clear from both Figure~\ref{fig:context}g and the right-hand column of Figure~\ref{fig:cavity_height} that the 
increased upward acceleration of the cavity occurred during the rise phase of the X-ray flare and as the nonthermal electron 
energy flux measured by Fermi approached its peak. This indicates that the upwards acceleration of the cavity was driven by the 
erupting flare below (consistent with the ``standard'' flare model) and suggests that the observed cavity was an erupting flux 
rope.

\begin{figure*}[!ht]
\centering
\includegraphics[width = 0.99\textwidth, trim=10 150 7 150,clip=]{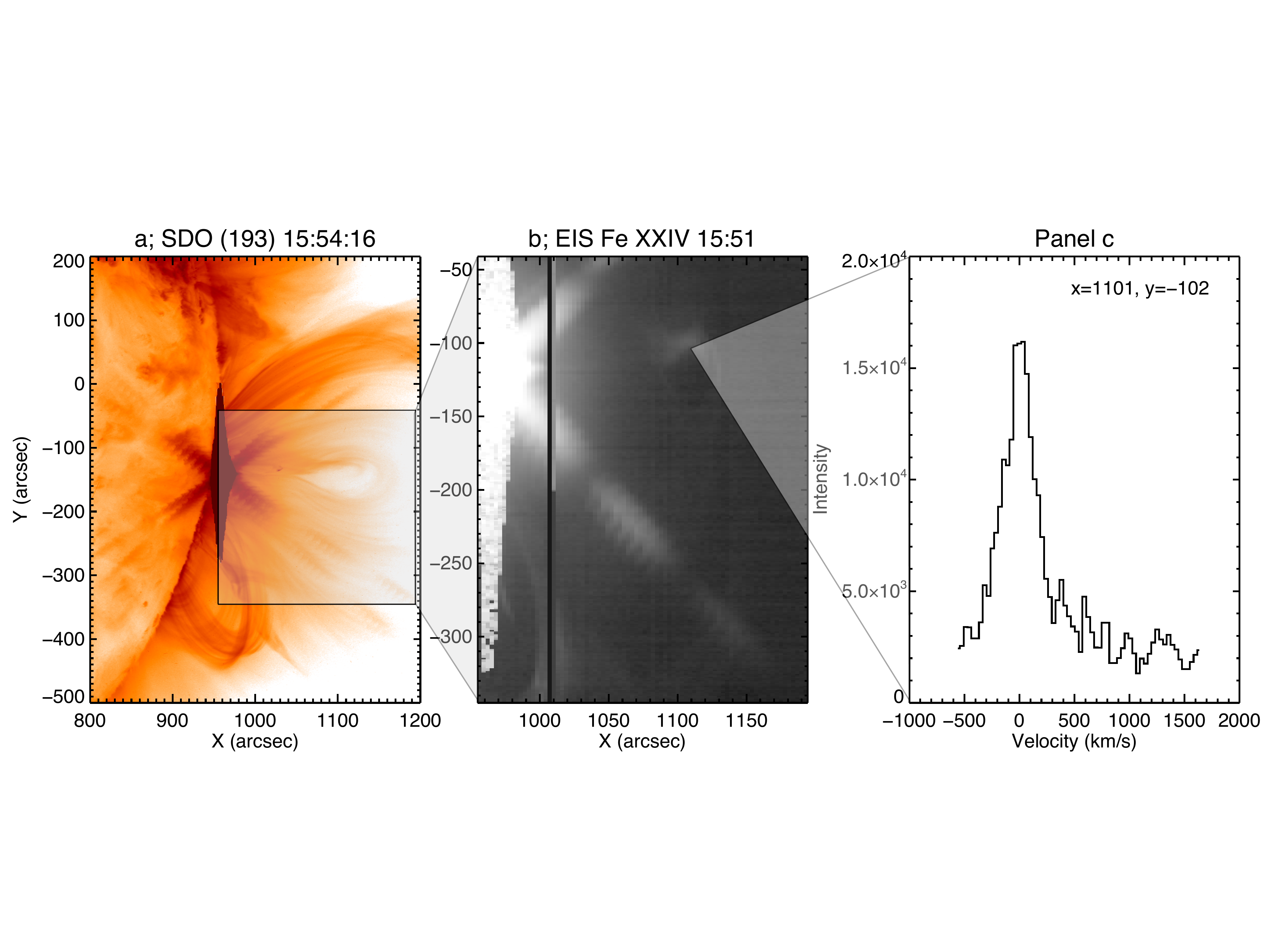}
\caption{\emph{Panel a}; SDO/AIA 193~\AA\ image at 15:54:16~UT using a reversed colour table and with the full EIS 
		 field-of-view highlighted. \emph{Panel~b}; Full field-of-view EIS Fe~\Rmnum{24} intensity image from the raster 
         beginning at 15:51~UT. \emph{Panel~c}; A sample Fe~\Rmnum{24} profile for the pixel at (1101, -102) as illustrated 
         in panel~b.}
\label{fig:EIS1551}
\end{figure*}

Although the cavity was best observed using the images from SDO/AIA, it was first observed by \emph{Hinode}/EIS. Due to 
its west-to-east rastering approach, EIS first observed the cavity between 15:48:00 and 15:49:20~UT (\ie\ during the raster 
which began at 15:42~UT) as it was first developing, and again at 15:53:20-15:54:40~UT as it erupted (during the raster which 
began at 15:51~UT). The cavity is most clearly seen in the Fe~\Rmnum{15} emission line shown in Figure~\ref{fig:EIS1542} for 
the raster beginning at 15:42~UT and the Fe~\Rmnum{24} emission line shown in Figure~\ref{fig:EIS1551} for the raster beginning 
at 15:51~UT. By comparison, SDO/AIA first observed the cavity at $\approx$15:50:00~UT (see Figure~\ref{fig:cavity_height}). 
Although the emission lines observed by EIS are also observed by SDO/AIA \citep[\cf][]{Odwyer:2010}, we believe that the 
SDO/AIA observations of the cavity were masked by the broadband nature of the passbands, with the result that it was not 
observed until the drop in intensity was sufficiently strong across the wavelength range of the given passband.


While no clear filamentary material can be observed in the Fe~\Rmnum{15} image at 15:42~UT shown in panel~b of 
Figure~\ref{fig:EIS1542}, strongly blue-shifted emission was observed in the region enclosed by the blue contours, 
corresponding to the bright filament emission observed in He~\Rmnum{2} shown in Figure~\ref{fig:EIS1542}c. The line spectra 
observed were incredibly complex, with an example of the blue-shifted emission shown in Figure~\ref{fig:EIS1542}d. While the 
Fe~\Rmnum{15} emission line is the dominant emission line in this portion of the spectrum under normal circumstances, the 
anomalous nature of the line spectrum shown in panel~d may suggest that this is not the case here. Other nearby lines which 
could be contributing include the nearby Al~\Rmnum{9} emission line at 284.03~\AA\ corresponding to a log~T=6.1, and a 
Fe~\Rmnum{17} emission line at 283.9~\AA\ corresponding to a log~T=6.9 \citep[\cf][]{Brown:2008,chianti:2013}. However, a 
comparable blue-shift to that seen in the Fe~\Rmnum{15} emission line was also observed in the Fe~\Rmnum{16} emission line, 
indicating that the blue-shift is indeed due to plasma motion rather than a contribution from the nearby Al~\Rmnum{9} and 
Fe~\Rmnum{17} emission lines. 

As noted above (and seen in the online movie associated with Figure~\ref{fig:context}), the filamentary material 
associated with the erupting cavity can be observed rising at the lower apex of the teardrop-shaped cavity in each SDO/AIA 
passband. It was also clearly observed in the EIS He~\Rmnum{2} data as a bright, very strongly blue-shifted blob (see 
Figure~\ref{fig:EIS1542}c and the spectrum in panel~e). In fact, the spectra in Figure~\ref{fig:EIS1542}d show that some of 
the spectra were nearly blue-shifted out of the spectral window, suggesting Doppler velocities greater than 200~\kms\ 
(consistent with the velocity of 214~\kms\ shown by the blue contours in panels b and c). As a cross-section of the cavity is 
observed here, this indicates that as the cavity expands and erupts, the plasma contained within it is flowing rapidly towards 
the observer, and most likely draining down the legs of the flux rope defined by the cavity. There is also some diffuse 
Fe~\Rmnum{24} emission around and across the cavity, consistent with hot, low density plasma in the core of the cavity.

The subsequent EIS raster began at 15:51:18~UT, with the corresponding Fe~\Rmnum{15} image shown in Figure~\ref{fig:EIS1551}b. 
At this time the cavity was observable in all AIA passbands, indicating a significant drop in density and/or temperature 
compared to the observations at the time of the earlier EIS raster shown in Figure~\ref{fig:EIS1542}. Although the intensity 
of the Fe~\Rmnum{15} emission line had dropped too much by this time to provide any usable observations, the Fe~\Rmnum{24} 
emission line (at Log~T=7.1) shows a bright edge around the cavity (shown in Figure~\ref{fig:EIS1551}b). However the 
corresponding spectra show that the strong plasma flow had mainly stopped by this time, with the spectra shown in 
Figure~\ref{fig:EIS1551}c showing a plasma velocity peaking at 0~\kms, with some slight broadening of the profile. This 
indicates that while the significant downward plasma flow associated with the eruption of the flux rope had mostly ceased 
by this time, there continued to be some draining of hot material from the cavity.

\section{Differential Emission Measure Analysis}
\label{s:dem}

\begin{figure*}[!ht]
\centering
\includegraphics[width = \textwidth, trim=0 0 0 0,clip=]{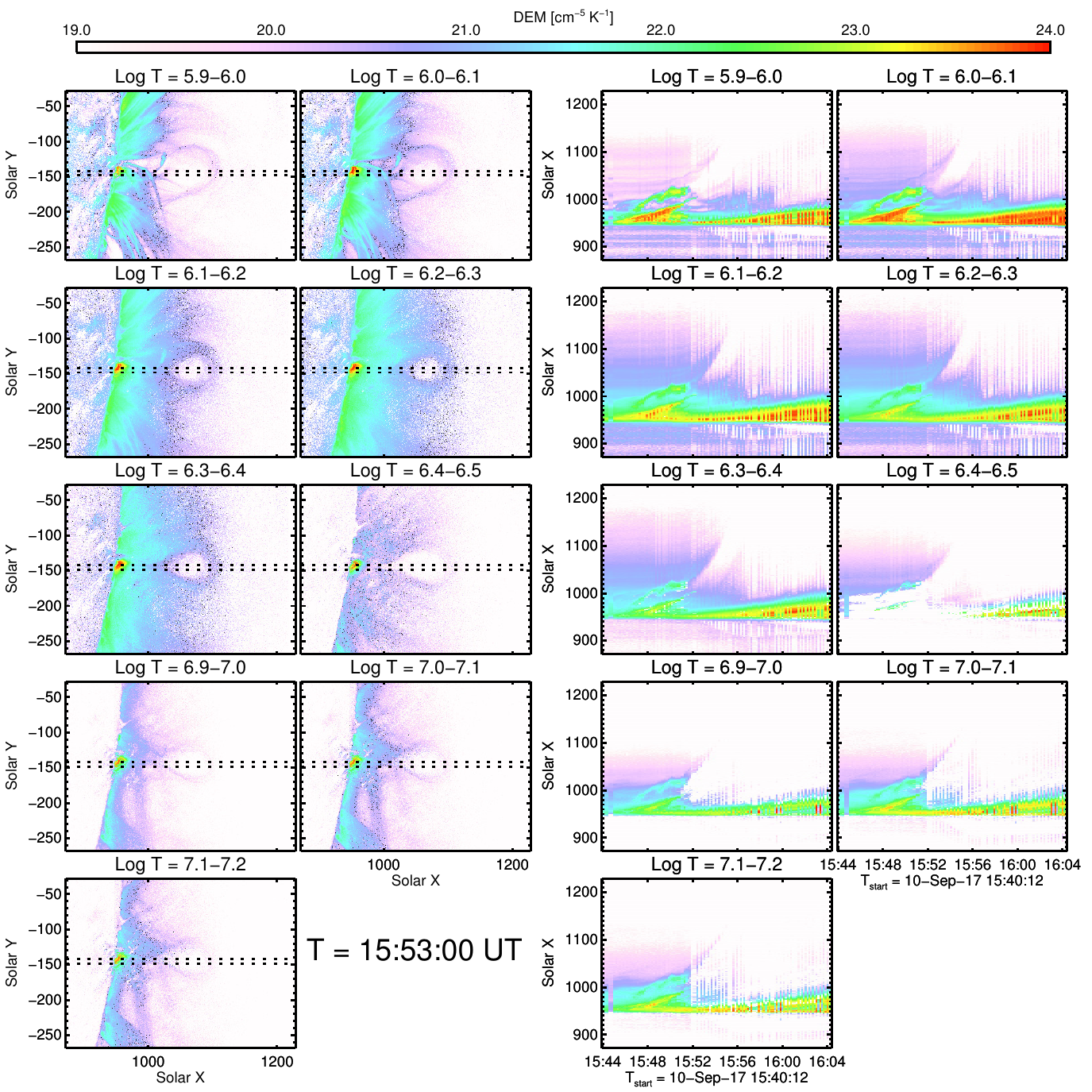}
\caption{Left two columns show the Differential Emission Measure (DEM) calculated from SDO/AIA EUV observations at 
		 $\sim$15:53:00~UT. Right two columns show stack plots of the temporal variation in DEM along the black dashed lines 
         shown in the left two columns. Although the DEM analysis produced 20 temperature bins of width Log~T=0.1 from 
         Log~T=5.7--7.7, the cavity is best seen in the temperature bins given here (\ie\ Log~T=5.9--6.5 and Log~T=6.9--7.2).}
\label{fig:DEM_plots}
\end{figure*}

While \emph{Hinode}/EIS gives a unique insight into the evolution and properties of the plasma during the eruption of the 
cavity, the very low temporal resolution complicates a complete analysis. The high temporal and spatial resolution data 
provided by SDO/AIA was therefore processed using the regularised inversion technique of \citet{Hannah:2013} to determine 
the differential emission measure (DEM) of the plasma within and surrounding the erupting cavity. This offers a complementary 
dataset which can be used to explore the properties and behaviour of the plasma at a much higher cadence. Although the Hannah 
\& Kontar technique was used to produce a DEM with 20 temperature bins of width Log~T=0.1 from Log~T=5.7--7.7, the cavity was 
best observed using the temperature bins from Log~T=5.9--6.5 and Log~T=6.9--7.2 (shown in the left two columns of 
Figure~\ref{fig:DEM_plots}). The clear observations of the erupting cavity in the range Log~T=5.9--6.5 match the cavity seen 
in the Fe~\Rmnum{15} emission line in Figure~\ref{fig:EIS1542}, while the edge of the cavity observed at Log~T=6.9--7.2 is 
consistent with the hot diffuse emission around the cavity observed in the Fe~\Rmnum{24} emission line at Log~T=7.1.

To investigate the temporal evolution of the filament plasma and the erupting cavity, a cut was taken along the black dashed 
line shown in each panel of the left two columns of Figure~\ref{fig:DEM_plots} and used to produce stack plots showing the 
temporal variation in DEM shown in the right two columns of Figure~\ref{fig:DEM_plots}. These stack plots show the gradual 
increase in the height of the filamentary material before it suddenly disappears in all temperature bins at $\sim$15:50:00~UT. 
At this point a cavity appears in most of the temperature bins before rising out of the field of view by $\sim$15:58:00~UT 
(consistent with the cavity evolution shown in the left-hand panel of Figure~\ref{fig:cavity_height}). These stack plots of 
differential emission measure also indicate that both the disappearing filamentary material and the surrounding envelope had a 
combination of cool and hot components.

\section{Discussion and Conclusions}
\label{sec:conc}

The combination of spectroscopic EIS and broadband AIA observations presented here describe the very fast eruption of a 
coronal cavity with a mixed temperature outer sheath and containing a combination of hot and cool plasma that exhibited very 
strong blue-shifts. This is a unique set of observations and as such requires some examination to determine the physical 
processes at work.

The temporal evolution of the coronal cavity obtained using the images from SDO/AIA indicates that the cavity \corr{initially 
began to rise slowly (with a velocity v$\sim$74--181~\kms). However, it then began to rise much more rapidly (at a velocity of 
v$\sim$439--513~\kms)}, with the Fermi non-thermal electron energy flux indicating that this was driven by continuous energy 
input from the flare below. The strongly blue-shifted plasma observed by \emph{Hinode}/EIS suggests that the rapid increase in 
height of the flux rope driven by the flare forced the plasma from the apex of the flux rope towards the legs, decreasing the 
density at the apex of the flux rope. As the flux rope \corr{initially} rose \corr{gradually}, this drop in plasma density would 
have been matched by a rise in plasma temperature, giving the observed combination of hot and cool plasma in the core of the flux 
rope. However, the \corr{subsequent} rapid rise of the flux rope and drop in its density also forced the expansion of the flux 
rope volume, producing the sudden appearance of the cavity in all wavelengths. As the cavity erupted and increased in height, the 
drop in pressure would have stopped the flow of plasma towards the legs of the flux rope, consistent with the lack of any clear 
Doppler motion in the Fe~\Rmnum{25} emission line.  

While the sudden off-loading of material from an erupting filament or flux rope has been previously observed 
\citep[\eg][]{Jenkins:2017}, this is the first time it has been observed spectroscopically for an eruption associated with an 
active region. These observations indicate that the eruption of the cavity was initially driven by the impulsive phase of the 
solar flare, with the increases in non-thermal electron energy matching the changing rise phase of the cavity. This rapid 
injection of a significant amount of energy forced a dramatic downflow of plasma from the apex of the erupting flux rope and 
meant that the erupting flux rope was much less dense and exhibited a different structure to other flux rope observations 
\citep[\cf][]{Hannah:2013}. These observations are consistent with the ``standard flare model'', and highlight the vital insight 
provided by spectrometers such as \emph{Hinode}/EIS.

\acknowledgments
\corr{The authors wish to thank the anonymous referee whose comments helped to improve the paper.} DML is an Early Career Fellow 
funded by the Leverhulme Trust. GAD and HPW were supported by NASA's Hinode project. JMJ thanks the STFC for support via funding 
given in his PhD Studentship. LKH and SAM acknowledge support from STFC via the Consolidated Grant ST/N000722/1. Hinode is a 
Japanese mission developed and launched by ISAS/JAXA, with NAOJ as domestic partner and NASA and STFC (UK) as international 
partners. It is operated by these agencies in co-operation with ESA and NSC (Norway). AIA data courtesy of NASA/SDO and the AIA, 
EVE, and HMI science teams. CHIANTI is a collaborative project involving George Mason University, the University of Michigan (USA) 
and the University of Cambridge (UK).



\end{document}